\documentclass{elsart}

\usepackage{natbib}

\usepackage{graphics}

\usepackage{amssymb}

\begin{document}

\begin{frontmatter}



\title{The Fundamental Plane of Quasars}


\author[affil1,affil2,affil3]{Timothy S. Hamilton, Stefano Casertano, and David A. Turnshek}
\address[affil1]{Shawnee State Univ., Dept. of Natural Sciences, 940 2nd St., Portsmouth, OH 45662; 
{\tt hamilton@milkyway.gsfc.nasa.gov}}
\address[affil2]{Space Telescope Science Institute}
\address[affil3]{University of Pittsburgh}

\begin{abstract}

We present results from an archival study of 70 medium-redshift QSOs
observed with the Wide Field Planetary Camera 2 on board the {\it
Hubble Space Telescope}.  The QSOs have magnitudes $M_V\leq -23$
(total nuclear plus host light) and redshifts $0.06\leq z\leq 0.46$.

A close relationship between QSO host and nucleus is found by examining multiple parameters
at once.  A principal components analysis shows that 3 nuclear and
host properties are related in a kind of fundamental plane: nuclear
luminosity and the size and effective surface magnitude of the bulge.
Using optical nuclear luminosity, this relationship explains 95.9\% of
the variance in the overall sample, while 94.9\% of the variance is accounted for 
if we use x-ray nuclear luminosity.

The form of this QSO fundamental plane shows similarities to the
well-studied fundamental plane of elliptical galaxies, and we examine
the possible relationship between them as well as the difficulties
involved in establishing this connection.  The key to the relationship
might lie in the fueling mechanism of the central black hole.

\end{abstract}

\begin{keyword}
quasars \sep 
galaxies: active
 

\PACS 98.54.-h

\end{keyword}

\end{frontmatter}


\section{Sample}
\label{sec:sample}

The sample is composed of 70 archival {\it Hubble Space Telescope}
({\it HST}) images of low-redshift QSOs.  They have redshifts between $0.06 \leq z \leq 0.46$ and total
(host plus nucleus) absolute magnitudes brighter than $M_V \leq -23$.
Furthermore, they must have been observed with the
{\it HST}'s Wide-Field Planetary Camera 2 (WFPC2), using broad-band
filters, and have images publicly available in the {\it HST} archives
as of 1999.  This brings our sample to 70 QSOs.
Rather than restrict our study to a specific class of QSOs, we impose
no physical criteria on the QSOs beyond those of magnitude and
redshift.  Thus we are able to study a broad range of properties and
draw general conclusions.  The images are reduced and the physical parameters fitted 
as described by Hamilton et al.~(2002).

\section{The ``Fundamental Plane'' of QSOs}
\label{sec:fp}

For our Principal Components Analysis (PCA), we use a restricted sample of those QSOs for which we
have all of the following parameters: $M_V\mathrm{(nuc)}$,  $L_X$, 
$r_{1/2}$, and $\mu_e$, where $\mu_e$ is the effective
surface magnitude of the galactic bulge.  We further
require that each QSO have a modeled, spheroidal bulge (the entire
galaxy, in the case of elliptical hosts).  These qualifications
restrict the sample to 42 QSOs.

We can perform two PCAs, an optical one using $M_V\mathrm{(nuc)}$,  
$\log r_{1/2}$, and $\mu_e$ as the parameters, and an x-ray one that 
substitutes $\log L_X$ for the nuclear luminosity.
From the optical PCA performed on this sample of 42 objects, 
we find that 96.1\% of the variance can be explained with just the first two
principal axes, and therefore the QSOs mostly lie in a plane within this parameter
space.  This we consider to be a fundamental plane (FP) for QSOs.  
For the corresponding x-ray results, the first two principal axes explain 
95.2\% of the variance in the sample, and we find here an x-ray QSO fundamental plane.
The individual subsamples of QSOs (radio-loud or radio-quiet, with spiral or elliptical hosts, and all 
combinations of these) 
are also examined in this way, and they show fundamental planes, as well.

\begin{figure}
\scalebox{0.675}{\includegraphics{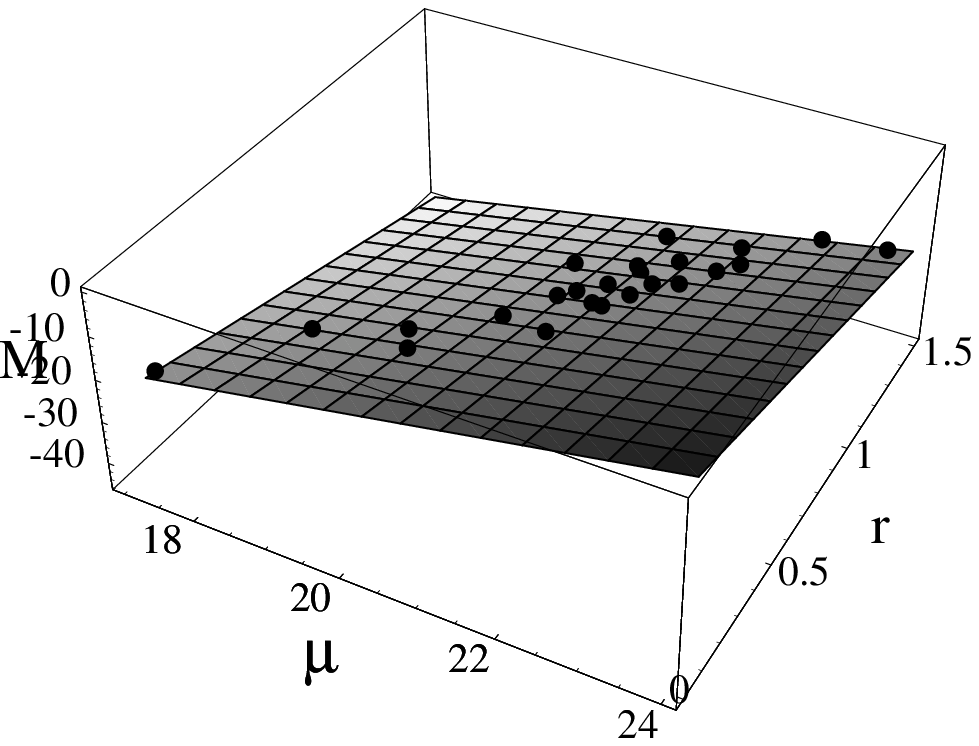}}
\scalebox{0.675}{\includegraphics{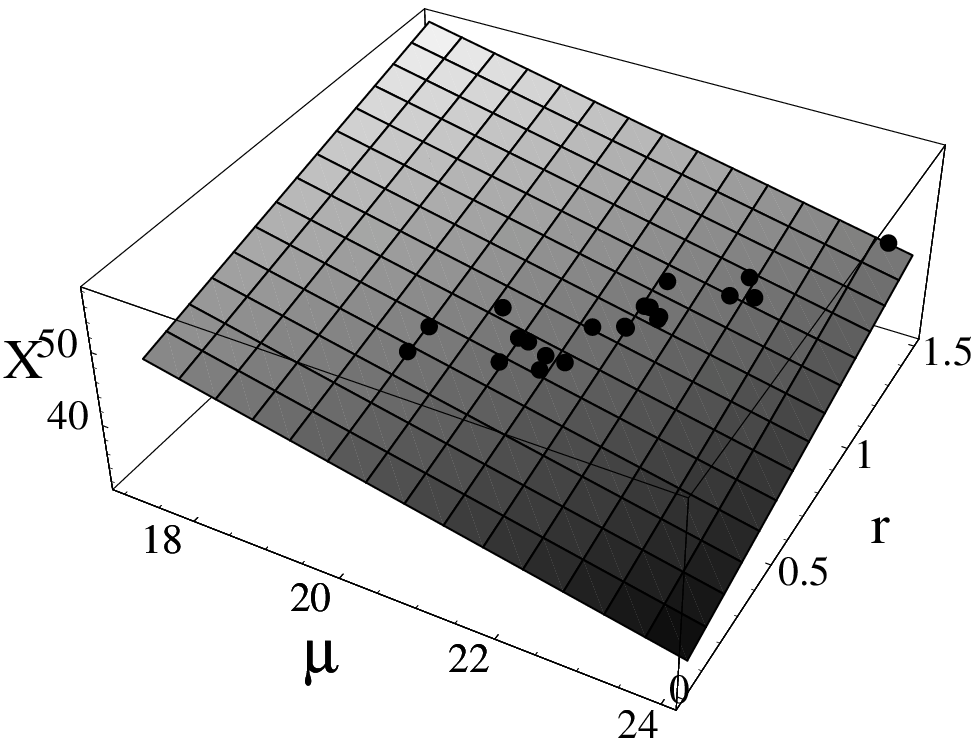}}
\caption{
Views of the optical (left) and x-ray (right) QSO
fundamental planes, showing the individual QSOs (points) and the plane fitted to the overall sample.  The host properties are the horizontal axes, while nuclear luminosity is vertical.   
The shading of the planes is proportional to nuclear luminosity, ranging from black (faint) to white (bright).  
Note that only those points lying above the plane are visible here.  
In the axis labels, ``M'' is $M_V\mathrm{(nuc)}$, ``X'' is $\log L_X$, 
``$\mu$'' is $\mu_e$, and ``r'' is $\log r_{1/2}$.
}
\label{fig:fp-phys}
\end{figure}

\begin{figure}
\scalebox{0.34}{\includegraphics{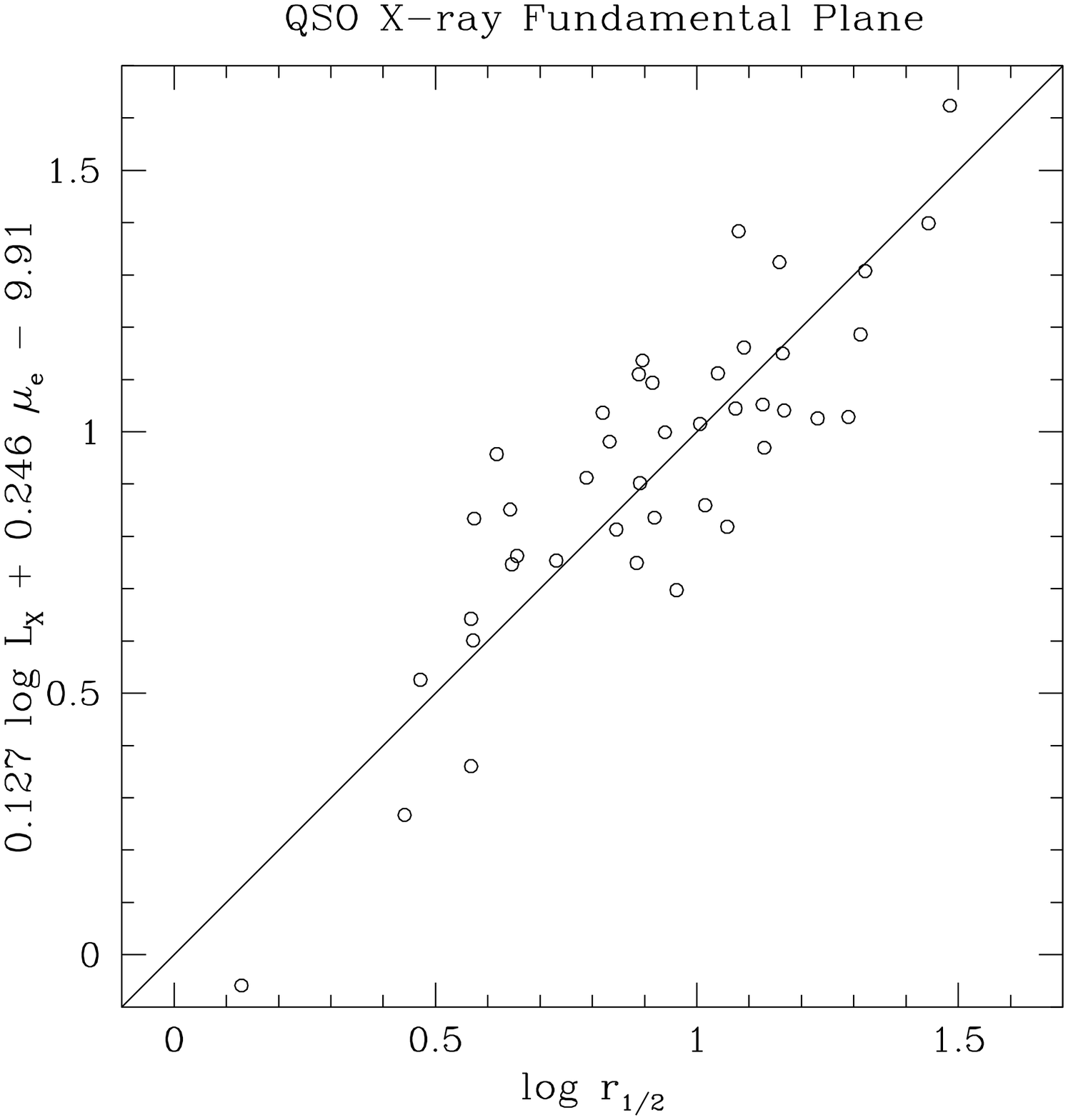}}
\scalebox{0.34}{\includegraphics{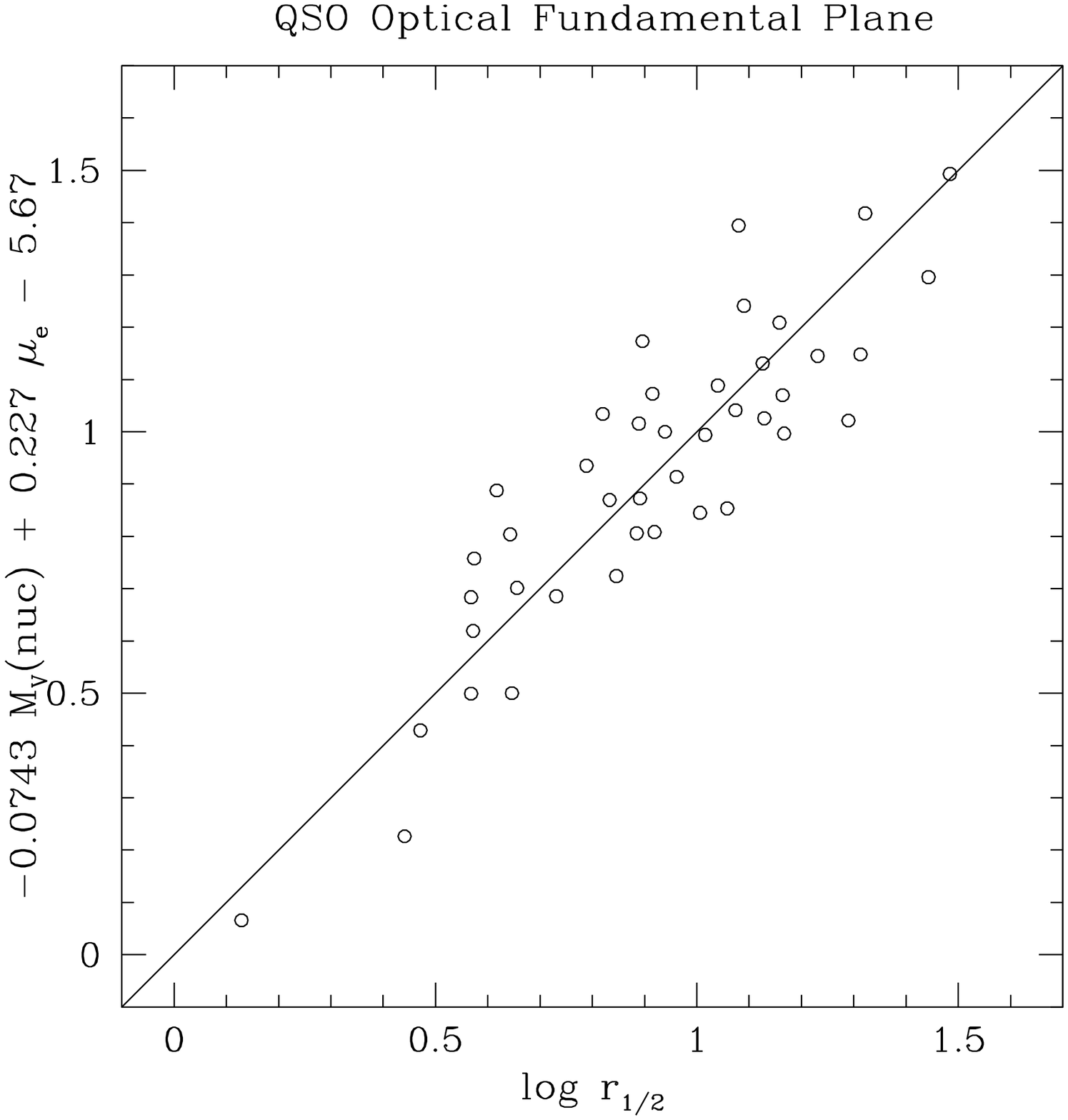}}
\caption{Overall QSO fundamental plane (vertical
axis), plotted against the measured host galaxy size ($\log r_{1/2}$, horizontal
axis).  Points on the diagonal line show perfect correspondence.  
The left figure uses the QSO fundamental plane in its optical form, while the right
figure uses the x-ray form.  The QSO fundamental plane is most precise when solved for the host size.}
\label{fig:fp-rms}
\end{figure}

We obtain the optical and x-ray formulae for the full sample's fundamental plane:
\begin{equation}
	M_V\mathrm{(nuc)} = -77.5 + 3.14 \mu_e - 14.2 \log
	r_{1/2}
	\label{equ:oall-physical}
\end{equation}
\begin{equation}
	\log L_X = 79.3 - 2.03 \mu_e + 8.74 \log
	r_{1/2} \mbox{ .} 
	\label{equ:xall-physical}
\end{equation}
Views of the optical and x-ray fundamental planes, with the QSO data points superimposed, 
are displayed in Figure~\ref{fig:fp-phys}.  Note that the host properties describe the horizontal and the nuclear luminosity the vertical in these plots.  Figure~\ref{fig:fp-rms} illustrates the precision of the 
QSO fundamental plane in both forms, with the plane plotted against 
the measured host sizes.  Its highest precision is found when solving for $\log r_{1/2}$.

\section{Discussion}
\label{sec:discussion}

\subsection{Possible Derivation}
The fundamental plane for QSOs shows a relationship between the
nuclear and host features that goes beyond the simple (and weak) correlation of nuclear and
host luminosities.  This behavior may be connected to other, known
relations between the objects.
For example, there is already a well-studied
fundamental plane for normal, elliptical galaxies (Djorgovski \&
Davis~1987; Dressler et al.~1987) that incorporates galaxy size, $r_{1/2}$, central velocity dispersion,
$\sigma_c$, and effective surface magnitude,
$\mu_e$.

Let us take a $V$-band measurement of the normal galaxy fundamental plane 
(Scodeggio et al.~1998), 
$\log r_{1/2} = 1.35 \log \sigma_c + 0.35 \mu_e + \mathit{Constant}$.  
The ratio of the coefficients of $\log r_{1/2}$ to $\mu_e$ differs by about 37\% between the QSO
optical fundamental plane and the normal galaxy FP, and the QSO x-ray FP
shows a 34\% difference.  Still, there is a formal similarity between
the QSO and normal fundamental planes, which might point to a link between 
the host galaxy's central velocity
dispersion and the nuclear luminosity of the QSO.  This
could derive from the fueling mechanism of QSOs, if the movement of gas
to the center of the galaxy and the black hole is related to the
velocity dispersion.

It is therefore tempting to try to derive the
QSO fundamental plane directly from the elliptical galaxy fundamental
plane, but we find two problems with this approach, both arising from the relation of black hole mass to nuclear luminosity.  Using the
velocity dispersion to black hole mass relation of Merritt \& Ferrarese~(2001),
$ \mathcal{M}_{BH}=1.3 \times 10^8 
	\left( \sigma_c / 200 \mbox{ km s}^{-1} \right)^{4.72} 
	\mathcal{M}_{\odot} $
we can put the elliptical galaxy fundamental plane in terms of black hole mass.  
Using the observed (but weak) correlation between black hole mass and 
nuclear luminosity in our sample, $M_V\mathrm{(nuc)} = -1.98 \log \left( \mathcal{M}_{\mathrm{BH}} / \mathcal{M}_{\odot} \right) -6.90$ and 
$\log L_X = 2.77 \log \left( \mathcal{M}_{\mathrm{BH}} / \mathcal{M}_{\odot} \right) + 19.8$,
we obtain
\begin{equation}
	M_V\mathrm{(nuc)} = \mathit{Constant} + 2.5 \mu_e - 7.14 \log r_{1/2}
\end{equation}
\begin{equation}
	\log L_X = \mathit{Constant} - 3.5 \mu_e + 10 \log r_{1/2} \mbox{ .}
\end{equation}
These are our attempts to derive the QSO optical fundamental plane from the normal galaxy FP.
The optical form differs from the actual QSO FP, equation~(\ref{equ:oall-physical}), 
completely outside the propagated errors, but 
the x-ray form is within the errors of equation~(\ref{equ:xall-physical}).

But any derivation of the QSO fundamental plane has an additional problem.  
As mentioned before, the QSO fundamental plane for the full sample is 
composed of individual FPs of the several subsamples.  Some subsample FPs 
actually slope in the opposite direction from the overall QSO FP.
For example, in the optical form, the FP of radio-quiets in elliptical hosts 
slopes in the opposite direction.  
And in the x-ray form, the radio-quiet subsamples slope oppositely 
from the overall sample.
Yet these differences cannot be accounted for by different correlations of 
nuclear luminosity with black hole mass.  
The poor correlation of black hole mass with nuclear luminosity lies in contrast with the 
relatively thin QSO fundamental plane.  
Furthermore, Woo \& Urry~(2002) suggest that the apparent correlations 
between black hole mass and nuclear luminosity are merely artifacts of 
sample selection.
Regardless of how we take this interpretation, the relationship between 
black hole mass and nuclear luminosity remains the missing link in any 
derivation of the QSO fundamental plane.

\subsection{Arrangement of Subsample Planes}

The insight into the origins of this new fundamental plane relationship might come from
the comparison of the QSO subsample FPs.  The thickness of the overall QSO
fundamental plane appears partly to be the result of the superposition of
the subsamples' planes.
Because the QSO FP mathematically describes a link between the host and the 
nucleus, it seems reasonable to suppose that the slope of the plane depends on the 
physical nature of this link.  The fueling mechanism at a QSO's core would seem to be the most directly related to this, depending on how we define ``fueling mechanism.''  
We could encompass within this term the details of the structure and dynamics of the accretion disk, as well as question of whether the QSO is efficiently or inefficiently fueled.

It is intriguing that as we change from one class to another, the fundamental plane 
essentially pivots about an axis, so the differences are mostly reduced to a single dimension, 
the slope (or gradient) relative to the $\mu_e$--$\log r_{1/2}$ plane.
The gradient directions, projected onto the $\mu_e$--$\log r_{1/2}$ plane, are almost all 
either aligned (or anti-aligned, for those with opposite slope).  
The optical subsample gradient directions are never more 
than 3.8 degrees away from that of the full sample, and in the x-ray form, they never exceed  
a 6.4 degree deviation.

Radio-loudness has the strongest effect on the slopes.  In the x-ray form, the subsample FPs 
are almost evenly divided between those aligned with the full sample and those anti-aligned.  
In the optical form, only radio-quiets in elliptical hosts tilt opposite to the full sample.
This effect is interesting because we are seeing a stark difference between the radio-loud and 
radio-quiet nuclei in the hosts of the same morphology.

It would be interesting to find if the different QSO FP orientations described above come 
about from different fueling mechanisms that might be found in the various subsamples.
We see, for instance, that radio-quiet and radio-loud QSOs are characterized by very 
different slopes in their x-ray FPs, but the understanding of what makes these QSO types differ 
is still too limited to speculate further here.  In our ongoing research, we are expanding the 
fundamental plane study to other types of AGN.  
We can then compare their FP orientations with those of the different QSO
subsamples, which may teach us more about the physics underlying the QSO fundamental plane.

\section{Future Work}

We should ask if lower luminosity classes of AGN (such as Seyferts or LLAGN) also have 
fundamental planes of this sort.  If they do, how do they compare with that of QSOs?  We can 
imagine four possibilities:
\begin{enumerate}
\item{They share the same fundamental plane as QSOs.  This would indicate that AGN power 
scales with the host properties, even across AGN types, and would support some form of 
unification.}

\item{The plane is parallel to that of QSOs, but shifted to lower nuclear luminosities.  This would 
show that these host properties don't determine the AGN class, and a given galaxy could 
host different types.}

\item{The plane is tilted with respect to that of QSOs.  Then the fundamental plane slope would 
be characteristic of the AGN type, possibly supporting the idea that the slope is tied to the 
accretion mechanism.}

\item{There is no fundamental plane whatsoever.  In that case, this type of fundamental plane 
would be a unique property of QSOs.  High-luminosity objects would be more closely 
connected with their host properties.}
\end{enumerate}

Any of these outcomes would teach us something useful.

\end{document}